\input harvmac
\noblackbox
\overfullrule=0pt
\def\Title#1#2{\rightline{#1}\ifx\answ\bigans\nopagenumbers\pageno0\vskip1in
\else\pageno1\vskip.8in\fi \centerline{\titlefont #2}\vskip .5in}

%
%

%
\def\IZ{\relax\ifmmode\mathchoice
{\hbox{\cmss Z\kern-.4em Z}}{\hbox{\cmss Z\kern-.4em Z}}
{\lower.9pt\hbox{\cmsss Z\kern-.4em Z}}
{\lower1.2pt\hbox{\cmsss Z\kern-.4em Z}}\else{\cmss Z\kern-.4em
Z}\fi}
\def\IB{\relax{\rm I\kern-.18em B}}
\def\IC{{\relax\hbox{$\inbar\kern-.3em{\rm C}$}}}
\def\ID{\relax{\rm I\kern-.18em D}}
\def\IE{\relax{\rm I\kern-.18em E}}
\def\IF{\relax{\rm I\kern-.18em F}}
\def\IG{\relax\hbox{$\inbar\kern-.3em{\rm G}$}}
\def\IGa{\relax\hbox{${\rm I}\kern-.18em\Gamma$}}
\def\IH{\relax{\rm I\kern-.18em H}}
\def\II{\relax{\rm I\kern-.18em I}}
\def\IK{\relax{\rm I\kern-.18em K}}
\def\IP{\relax{\rm I\kern-.18em P}}
\def\IR{\relax{\rm I\kern-.18em R}}

%

%
\def\II{\relax{I\kern-.10em I}}

\def\IIb{{\II}b}
\def\e{\epsilon}
\def\apm{{\alpha^\prime}}

\def\[{\left [}
\def\]{\right ]}
\def\({\left (}
\def\){\right )}
\def\p{\partial}
\def\vol{{\rm vol~}}
\lref\dasmathurtwo{S. Das and S. Mathur,
``{\it Interactions involving D-branes}'',
hep-th/9607149.}
\lref\msg{J. Maldacena and A. Strominger, hep-th/9609026.}
\def\s{\sigma}
\def\e{\epsilon}
\def\[{\left [}
\def\]{\right ]}
\def\p{\partial}
\def\({\left (}
\def\){\right )}
\lref\spin{R. Breckenridge, R. Myers, A. Peet and C. Vafa,
``D--branes and Spinning Black Holes,'' hep-th/9602065.}
\lref\blmpsv{R. Breckenridge, D. Lowe, R. Myers, A. Peet, A.
Strominger
and C. Vafa, ``Macroscopic and Microscopic Entropy of Near-Extremal Spinning
Black Holes,'' hep-th/9603078.}
\lref\pton{C. Callan, S. Gubser, I. Klebanov and A. Tseytlin,
hep-th/9610172.}
\lref\klma{I. Klebanov and M. Krasnitz, hep-th/9612051.}
\lref\gk{ S. Gubser and I. Klebanov,
``Emission of Charged
particles
from four- and five-dimensional black holes'',
hep-th/9608108.}
\lref\page{ D. Page, Phys. Rev. {\bf D 13} (1976) 198; Phys. Rev. {\bf
D14}
(1976), 3260.}
\lref\unruh{W. Unruh,
Phys. Rev. {\bf D14} (1976) 3251.}
\lref\cj{ M. Cvetic and D. Youm, hep-th/9512127.}
\lref\hms{G. Horowitz, J. Maldacena and A. Strominger,
``Nonextremal Black Hole Microstates and U-duality'',
 hep-th/9603109.}
\lref\dbr{J. Polchinski, ``TASI Lectures on D-Branes,'' hep-th/9611050.}
\lref\shenker{S. H. Shenker, ``Another Length Scale in String Theory?'',
hep-th/9509132.}
\lref\jp{J. Polchinski, Phys. Rev. Lett. 75 (1995) 4724--4727, hep-th/9510017.}
\lref\dm{S. Das and S. Mathur, ``
Comparing decay rates for black holes and D-branes'',hep-th/9601152.}
\lref\ghas{G. Horowitz and A. Strominger,``Counting States of
Near-extremal Black Holes'',  hep-th/9602051.}
\lref\ascv{A. Strominger and C. Vafa, Phys. Lett. {\bf B379} (1996)
99,
 hep-th/9601029.}
\lref\hrva{P. Horava, Phys. Lett. {\bf B231} (1989) 251.}
\lref\cakl{C. Callan and I. Klebanov, hep-th/9511173.}
\lref\prskll{J. Preskill, P. Schwarz, A. Shapere, S. Trivedi and
F. Wilczek, Mod. Phys. Lett. {\bf A6} (1991) 2353. }
\lref\bhole{G. Horowitz and A. Strominger,
Nucl. Phys. {\bf B360} (1991) 197.}
\lref\bekb{J. Bekenstein, Phys. Rev {\bf D12} (1975) 3077.}
\lref\hawkirr{S. Hawking, Phys. Rev {\bf D13} (1976) 191.}
\lref\stas{A.~Strominger and S.~Trivedi,  Phys.~Rev. {\bf D48}
 (1993) 5778.}
\lref\bek{J. Bekenstein, Lett. Nuov. Cimento {\bf 4} (1972) 737,
Phys. Rev. {\bf D7} (1973) 2333, Phys. Rev. {\bf D9} (1974) 3292.}
\lref\hawkb{S. Hawking, Nature {\bf 248} (1974) 30, Comm. Math. Phys.
{\bf 43} (1975) 199.}
\lref\cm{C. Callan and J. Maldacena, Nucl. Phys. {\bf B 475 } (1996)645,
hep-th/9602043.}
\lref\tseytlin{A. Tseytlin, hep-th/9601119.}
\lref\tata{S. Dhar, G. Mandal and S. Wadia, ``
Absorption vs Decay of Black Holes in String Theory and
T-symmetry'', hep-th/9605234.}
\lref\gm{G. Horowitz and D. Marolf, hep-th/9605224.}
\lref\mss{J. Maldacena and L. Susskind, ``D--branes and Fat
Black Holes'',  hep-th/9604042.}
\lref\bl{V. Balasubramanian and F. Larsen, hep-th/9604189.}
\lref\jmn{J. Maldacena,``Statistical Entropy of Near Extremal
five-branes'', hep-th/9605016.}
\lref\hlm{G. Horowitz, D. Lowe and J. Maldacena, hep-th/9603195.}
\lref\hgm{J. Gauntlett, J. Harvey, M. Robinson and D. Waldram,
Nucl. Phys. B411 (1994) 461--472, hep-th/9305066.}
\lref\cmp{C. Callan, J. Maldacena and A. Peet,
Nucl. Phys. B475 (1996) 645, hep-th/9510134.}
\lref\hs{G. Horowitz and A. Strominger}
\lref\hms{G. Horowitz, J. Maldacena and A. Strominger,
hep-th/9603109.}
\lref\chs{C. Callan, J. Harvey and A. Strominger,
Nucl. Phys. B367 (1991) 60--82.}
\lref\gf{M. R. Douglas, ``Gauge Fields and D-branes,''
hep-th/9604198.}
\lref\samstr{T. Samols and A. Strominger, unpublished notes (1993).}
\lref\kp{D. Kabat and P. Pouliot,``A
Comment on Zero-Brane Quantum Mechanics", hep-th/9603127.}
\lref\dkps{M. R. Douglas, D. Kabat, P. Pouliot and S. Shenker,
``D-Branes and Short Distances in String Theory", hep-th/9608024.}
\lref\fks{S. Ferrara, R. Kallosh and A. Strominger,
Phys. Rev. D52 (1995) 5412--5416, hep-th/9508072.}
\lref\cmpf{
C.G. Callan, E.J. Martinec, M.J. Perry and D. Friedan,
Nucl. Phys. B262 (1985) 593.}
\lref\sdd{See Section 7 of
{\it Supergravities in Diverse Dimensions},
eds. A. Salam and E. Sezgin,
North Holland/World Scientific (1989).}
\lref\dougrev{M. R. Douglas, Nucl. Phys. (Proc Suppl) 41, 66.}
\lref\ggo{D. J. Gross and R. Gopakumar, Nucl. Phys. B451 (1995) 379.}

\Title{\vbox{\baselineskip12pt\hbox{hep-th/9703031}}}
{\vbox{
\centerline{Probing Five-Dimensional Black Holes} 
\centerline{with D-Branes}}}
\centerline{Michael Douglas}
\vskip 0.1in
\centerline{\it Department of Physics and Astronomy, Rutgers University,
Piscataway, NJ}
\vskip 0.1in
\centerline{Joseph Polchinski}
\vskip 0.1in
\centerline{\it Institute for Theoretical Physics, University of California,
Santa Barbara, CA }
\vskip 0.1in
\centerline{\it and}
\vskip 0.1in
\centerline{Andrew Strominger}
\vskip 0.1in
\centerline{\it Department of Physics, University of California,
Santa Barbara, CA}
\vskip 0.2in
\centerline{\bf Abstract}
\vskip 0.3in
{We consider a one-brane probe in the presence of a 
five-dimensional black hole in the classical limit. The velocity-dependent
force on a slowly-moving probe is characterized by a metric on 
the probe moduli space. 
This metric is computed for large black holes using low-energy 
supergravity, and for 
small black holes using D-brane gauge theory. The results are compared.}
\Date{}
\newsec{Introduction}

Five-dimensional black holes in string theory \ascv\ are primarily
characterized by two quantities. The first is ${gQ\apm}$,
where $g$ is the closed-string coupling and $Q$ is the number of
minimal units of charge or constituent D-branes \jp.
This is a measure of the size of the black hole. The second is $g$
itself which governs the strength of quantum corrections.

In this paper we shall primarily consider the classical limit
$g \to 0, ~Q \to \infty$  with the black hole size
$gQ$ held fixed in string units. For $g=0$  there is no
Hawking radiation. Within this limit there are
two distinct regimes. Large black holes have $gQ\gg 1$ and are well
described by general relativity or, more precisely, classical closed
string perturbation theory. Small black holes have $gQ\ll 1$ and so are
smaller than the size of a typical string. They are well
described by D-brane perturbation theory \jp. The adjective
``black'' is in all cases
appropriate for these objects\foot{One might question the use of the word ``hole''
for the small objects. However we shall later see that the probe moduli space 
metric has a hole in it.} because,
as pointed out in \refs{\hawkirr,\tata}, the formula for their entropy
implies that light cannot be emitted regardless of their
size for $g\to 0$. 

There is by now overwhelming evidence that for very
low energy processes there is a single effective description
-- the effective string \refs{\ascv,\ghas,\mss}\ -- which is valid under some
circumstances (the dilute gas regime \refs{\ghas,\hms})
for both large and small black holes. This effective description
correctly yields the extremal \refs{\ascv,\spin}
and near-extremal \refs{\ghas,\blmpsv}
entropies as well as the total Hawking radiation
rate \refs{\cm,\dm}
and its functional dependence on various parameters \refs{\msg,\pton,\klma}.
Hence for low energy processes there is no qualitative difference
between large and small black holes.

However a light, low energy probe misses a defining feature of
a black hole: the event horizon.
Such a probe is necessarily large and cannot fully fall through the event
horizon. In order to directly see the event horizon of a small
black hole, we need a probe that is smaller than the string length.
In light of \shenker\ D-branes are natural candidates and
a series of works starting with \refs{\kp,\gf}\ found
that D-branes are quite effective at probing distances shorter
than the string length.
General arguments imply that this regime is well-described by D-brane world-volume
gauge theory and numerous examples are given in \dkps.

Accordingly, in this paper we use D-branes to
probe the structure of black holes. Specifically, we consider the
moduli space metric for a wrapped one-brane probe in the presence
of the five-dimensional black hole of \ascv.
Our results are puzzling and inconclusive.
For large black holes
the moduli space metric is computed using a low energy spacetime-probe
action.  While the coefficients of the string metric are infinite series
in $1/r$, we find that the probe moduli space metric
has only $1/r^2$ and $1/r^4$ corrections. The $1/r^2$ piece
follows simply from the long-range force laws. The $1/r^4$
term dominates the near-horizon behavior
and has a universal, moduli-independent coefficient, reflecting the
fact that near-horizon geometry forgets the asymptotic
moduli\fks.
Next we compare this to the D-brane calculation
valid for small black holes.  We expected -- based
on previous examples -- that the D-brane calculation would produce the
same metric.  We indeed find that the $1/r^2$ term
with the correct coefficient arises from a one-loop calculation
in the D-brane gauge theory.  However, although various considerations
suggest that $1/r^4$ term should arise at two loops, we 
were unable to find it. Possibly we missed a subtlety in the 
two-loop calculation, as discussed at the end of Section 4.
It is
our view that resolution of this issue is crucial to further progress in this
general direction.

This paper is organized as follows.  Section 2 contains the basic results
from the low energy classical closed string point of view -- the
derivation and justification of the probe action, and the demonstration that
a probe approaching the black hole will be captured.
We also consider general constraints on the probe metric, including those
following from U-duality, and the behavior of a probe inside the event horizon.
Section 3 extends this to near-extremal black holes, which exert a force 
on static probes.
Section 4 analyzes the same system from the D-brane point of view.
We derive the effective world-volume gauge theory on the probe and
compute quantum corrections.  Section 5 develops the constraints
imposed by supersymmetery on the probe metric.

\newsec{Slowly Moving
One-branes and Five-branes in a Black Hole Background}

\subsec{The Extremal Black Hole Solution}

The low-energy action for ten-dimensional type \IIb\ string theory
(in the notation of \hms) contains the terms,
\eqn\fds{{1\over 16 \pi G_{10}}
\int d^{10}x \sqrt{- g}e \[ e^{-2\phi}R+4e^{-2\phi}(\nabla \phi )^2
-{1 \over 12} H^2 \]}
in the ten-dimensional string frame.
$H$ denotes the RR three form field
strength, and $\phi$ is the dilaton. The NS three form, self-dual five
form, and second scalar are set to zero.
We leave $\alpha'$ explicit in this section, but 
set it to one in Section 3.
We let $g$ denote
ten-dimensional string coupling and define the zero mode of
$\phi$ so that
$\phi$ vanishes asymptotically. The ten-dimensional Newton's constant is then
$G_{10}=8 \pi^6 g^2 {\apm}^4$.
We wish to consider toroidal compactification to
five dimensions with an $S^1$ of length $2\pi R$ and
a $T^4$ of four-volume $(2\pi)^4 V$.
With
these conventions, T-duality sends $R$ to $\apm/R$ or $V$ to $\apm^4/V$, and
S-duality
sends $g$ to $1/g$.

A  ten-dimensional extremal solution labeled by three charges is given by
\eqn\dil{ e^{-2\phi } = {Z_5 \over Z_1},}
\eqn\htn{H= 2 r_5^2 \e_3 + 2 r_1^2  e^{-2\phi}  *_6 \e_3,}
\eqn\metric{\eqalign{
ds^2 = &Z_1^{-1/2}
Z_5^{-1/2}
  \[ - dt^2 +dx_5^2+{r_n^2 \over r^2} ( dt +  dx_5)^2\]
\cr
+&
 Z_1^{1/2}
Z_5^{-1/2}dx^m dx^m  +Z_1^{1/2} Z_5^{1/2}\left[
 dr^2 + r^2 d \Omega_3^2 \right]
,}}
\eqn\dfr{\eqalign{Z_1&\equiv 1+{r_1^2 \over r^2}
\qquad r_1^2 \equiv {gQ_1\apm^3 \over V},\cr
Z_5&\equiv 1+{r_5^2 \over r^2} \qquad  r_5^2 \equiv gQ_5\apm,\cr
Z_n&\equiv 1 + {r_n^2 \over r^2} \qquad r_n^2 \equiv { g^2n \apm^4\over R^2V},
}}
where $*_6$ is the Hodge dual in the six dimensions $x^0,..,x^5$
and $\e_3$ here is the volume element on the unit three-sphere.
The event horizon is at $r=0$.
$x^5$ is periodically identified with period $2\pi R$,
$x^m$,  $m = 6,...,9$, are each identified with period $2\pi V^{1/4}$.
The three charges are defined by
\eqn\charges{
\eqalign{
   Q_1 &= {V\over 4\pi^2 g\apm^3}\int e^{2\phi} *_6H , \cr
   Q_5 &= {1\over 4\pi^2 g\apm} \int H ,
\cr
  n &= {RP} ,
}}
where $P$ is the total momentum around the $S^1$.
All charges are normalized to be integers and taken to be positive.

The entropy and energy are 
\eqn\extl{\eqalign{E &={1 \over g^2}\left[  {R gQ_1\over \apm}  +
{  R V gQ_5\over \apm^3 }  +
 { g^2 n  \over R}\right]  ~,\cr
S&=2\pi\sqrt{Q_1Q_5n}~~.}}

\subsec{The Classical Limit}

In the classical limit the action becomes very large and
the stationary phase approximation can be applied. Since the
action \fds\ has an explicit $1/g^2$ prefactor, the limit
$g\to 0$ with the fields held fixed is a classical limit.
Noting the explicit factors of $1/g$ in the definitions
\charges\ of the integer charges, as well as the explicit
$1/g^2$ in the definition of the energy $E$ and momentum $P$,
this is equivalent to
\eqn\clim{g \to 0 \qquad
{\rm with} ~~gQ_1, ~~gQ_5, ~~g^2 n ~~{\rm fixed.}}
While the quantized charges diverge in the limit \clim,
the classical solutions remain finite.
The canonical energy and momentum
come with explicit $1/g^2$ factors, but this divergence is conventional
and could be eliminated by using units in which $\hbar\ne 1$.
However, the $1/g^2$ divergence of the entropy is meaningful.

Closed string perturbation theory naturally treats the fields
$\phi,~~g$ and $H$ as order
one. Hence, noting the explicit factors of $1/g$ in \charges,
it is an expansion in $g^2$ with $gQ_1, ~~gQ_5$
and $g^2 n$ fixed. Thus the classical limit \clim\ corresponds to
genus zero closed string theory.  A primary tool for
analyzing black hole solutions in classical closed string theory
is the $\apm$ expansion. The solutions \dil -\metric\ are solutions of
the leading order $\apm$ equations. They are characterized by
the squared length scales $gQ_1\apm , ~~gQ_5\apm $
and  $ ~~g^2 n\apm $ -- in particular, the curvatures are bounded by
these scales at the horizon.  Thus,
when these are large
in string units:
\eqn\cspert{gQ_1>1, ~~gQ_5>1,~~g^2n>1,}
the $\apm$ expansion is valid everywhere outside the horizon.

In section 4 we will compare this solution with the D-brane realization of
the same black hole.  D-brane perturbation theory involves
both open and closed string
loops. Closed string loops have factors of $g^2$, while open string
loops have factors of $gQ_1$ or $gQ_5$, because
the open string loops can end on any of the D-branes. Hence the
classical limit \clim\ is a large $N$ limit
of the open string field theory.
Closed string loops are suppressed, and
the large $N$ limit is the sum over planar open string diagrams with
arbitrarily many boundaries.

This open string theory also admits an $\apm$ expansion, but now in
powers of $r^2/\apm$, where $r$ is a separation between D-branes,
the parameter controlling the mass of the lightest open strings stretched
between D-branes.  When
\eqn\oplight{r^2 \ll  \apm,}
the qualitative physics and
in particular the leading singular behavior of the theory is given by
a world-volume quantum field theory keeping only these modes,
while excited open string and closed string effects only make non-singular
corrections \dkps.  Thus the near-horizon behavior should be described by
the large $N$ limit of a quantum field theory\foot{Assuming, as we believe to be the 
case, that $r=0$ in D-brane coordinates indeed coresponds to the horizon.}.

Note that \cspert\ and \oplight\ are conditions on different quantities
and thus have a simultaneous regime of validity, the near-horizon behavior
of a large black hole.  Hence in principle large N 
D-brane gauge theory can be used to 
study the event horizon!
Unfortunately, at present our ability
to study the large $N$ limit is limited.
Most of what we know comes from taking the large $N$ limit of exact
results at finite $N$.
The main result which is more general than this
is large $N$ factorization, which states that correlators of gauge
invariant operators $\vev{\prod_i \Tr O_i}$ are dominated by the
disconnected part $\prod_i\vev{ \Tr O_i}(1+O(1/N))$.
This translates into a precise sense in which the black hole is classical --
if we express coefficients in the probe action in terms of gauge invariant
operators formed from the black hole degrees of freedom, by evaluating
the operators in the black hole configuration, we derive an action for
probe motion in a constant background.\footnote*{One can also phrase
this as the existence of a ``master field'' which determines gauge
invariant quantities; see \refs{\dougrev,\ggo} for an introduction.
A master field exists which reproduces any set of gauge invariant
correlation functions.}

For now, we
will have to work with conventional open string perturbation theory.
This is good if
\eqn\dpert{gQ_1<1, ~~gQ_5<1, ~~g^2n<1}
and the regimes are mutually exclusive.
We shall see that the last condition arises because a
correlation function can pick up a factor of $g^2 n$ via
propagators hooking on to the external state.

To summarize, the classical limit \clim\ may be characterized either
by the classical genus zero closed string theory or by the large
$N$ limit of the quantum D-brane open string theory.
These two different representations of the
limit \clim\ truncate to field theory (and thus are useful)
in different regimes of the couplings
according to \cspert\ and \dpert. The Lagrangian \fds\ is good
for black holes large compared to the string scale, while the
D-brane quantum field theory is good for small black holes.

\subsec{One-brane and Five-brane probes}

Consider a slowly moving D one-brane wound around the
$S^1$ in the black hole background
\dil - \metric. It is convenient to choose static
gauge $\tau=t,~~\sigma =x^5$. The one-brane is then described by the ansatz
\eqn\oneans{\eqalign{X^0&=\tau , \cr X^5&=-\sigma, \cr
                  X^m&=X^m(\tau), \cr X^i &=X^i(\tau), \cr }}
where $i,j=1,2,3,4$ are indices in then non-compact
transverse space and $m,n=6,7,8,9$ are internal $T^4$ indices.
The one-brane worldsheet action, obtained for example by S-duality from
the fundamental string action, is
\eqn\onews{-
{1\over 2 \pi g\apm}\int d\tau d\sigma\ e^{-\phi}\sqrt{-\det g_{\mu\nu}\p X^\mu \p X^\nu}
+{1\over 2 \pi g\apm}\int C^{(2)},
}
where $C^{(2)}$ is the potential obeying $H=dC^{(2)}$and $\mu,\nu=0,...9$.
The action governing the
dynamics of a slowly moving one-brane in the gauge \oneans\
is then obtained by inserting the
ansatz \oneans\ into this action. This yields
\eqn\onemod{ -{R\over g \apm}
\int d\tau  +
{R\over 2 g\apm }
\int d\tau \( Z_nZ_5 v^2+Z_nw^2 \) +{\cal O}(v^4),}
where $v^2\equiv \dot X^i\dot X^i$ is the squared transverse velocity
and $w^2\equiv \dot X^m\dot X^m$ is the squared velocity in the $T^4$.
The first term in \onemod\ represents the action of a one-brane at
rest.
In deriving this action we have used a cancellation
(required by supersymmetry) between contributions from
$g_{00}, ~~g_{55}$ and the two form potential $C_{05}$ related
to \htn.

The motion of the one-brane is hence a geodesic in the wormhole
geometry
\eqn\wrone{ds_{M1}^2=Z_nZ_5\left[
dr^2 + r^2 d \Omega_3^2 \right]+Z_n dx^mdx^m.}
Near $r=0$ the first term approaches the flat metric on $R^4$
\eqn\wroneflat{ds_{M1}^2 \to { g^3nQ_5 \apm^3\over R^2V}\left[
d\rho^2 + \rho^2 d \Omega_3^2 \right],}
where $\rho=\apm/r$. Hence $r=0$ is a second asymptotic region.

For the special case of $n=0$,
the metric \wrone\ has appeared previously in the literature
\chs\ as the transverse string metric of the symmetric five-brane
carrying NS charge. This is not a coincidence.
Under S-duality  $Q_5$ becomes the NS charge and
the probe becomes a fundamental string.
In this context it was argued that $(4,4)$ nonrenormalization
theorems protect the metric from corrections. We expect the same to
hold in the present context. For the special case of $Q_5=0$,
\wrone\ is S-dual to the metric for scattering of
Dabholkar-Harvey winding states which appeared in \refs{\hgm,\cmp}.
These two special cases are related to one another by a U-duality
transformation which exchanges $n$ with $Q_5$.

Next let us consider the five-brane, with world-volume action proportional to
\eqn\fivewv{-{1 \over g {\apm}^3}
\int d\tau d^5\sigma\ e^{-\phi}\sqrt{-\det g_{\mu\nu}\p X^\mu \p X^\nu}
+ {1 \over g}\int  C^{(6)} .
}
Inserting an ansatz similar to \oneans\ but with $X^i$ wrapping
the $T^4$ produces the same metric multiplied by the factor
\eqn\fivefactor{
e^{-2\phi }\vol^2 (T^4) = {Z_1 \over Z_5}.}
Now the $r$-dependent terms in $g_{00}$
and $g_{ij}$ cancel against the six-form potential
$C^{(6)}$
leaving the metric on the transverse moduli space
\eqn\wrfve{ds_{M5}^2=Z_nZ_1\left[
dr^2 + r^2 d \Omega_3^2 \right].}
Note that the transverse part of \wrone\ and \wrfve\ are exchanged under
$T$-duality
of the internal $T^4$, as expected. One could also compute a metric for slow
variations of the Wilson lines in the five-brane which would be T-dual to the
second term in \wrone.

The simplicity of \wrone\ and \wrfve\ is quite striking.
While the black hole metric itself contains highly nonlinear
corrections, these moduli space metrics truncate after
$1/r^4$ corrections. This corresponds to at most
one and two loops in the D-brane perturbation expansion.

\subsec{Inside The Event Horizon}

The black hole solution \metric\ has an event
horizon at the  coordinate singularity $r=0$.
The region inside the event horizon is described by the geometry
\nref\homa{G. Horowitz and D. Marolf, hep-th/9610171.}
 \eqn\idil{ e^{-2\phi } = {Z_5 \over Z_1},}
\eqn\ihtn{H= 2 r_5^2 \e_3 -2 r_1^2  e^{-2\phi}  *_6 \e_3,}
\eqn\imetric{\eqalign{
ds^2 = &Z_1^{-1/2}
Z_5^{-1/2}
  \[  dt^2 -dx_5^2+{r_n^2 \over r^2} ( dt +  dx_5)^2\]
\cr
+&
 Z_1^{1/2}
Z_5^{-1/2}dx^m dx^m  +Z_1^{1/2} Z_5^{1/2}\left[
 dr^2 + r^2 d \Omega_3^2 \right]
,}}
where in the interior region
\eqn\idfr{\eqalign{Z_1&\equiv -1+{r_1^2\over r^2},\cr
Z_5&\equiv -1+{r_5^2 \over r^2}.\cr}}
This differs from the exterior geometry only in the sign
of $dt^2-dx_5^2$, signs in the definitions of
$Z_1,~Z_5$, and a sign in the second term in $H$. This
last sign arises because $dr$ is inward-pointing in the
interior region (the horizon is still at $r=0$). It has the
interesting consequence \homa\ that only negatively charged,
anti-one-branes are static inside the horizon.\foot{
These have locally positive energy, but their energy as measured from
spatial infinity in
the exterior region is negative.
Hence the black hole + probe configuration can still
saturate the BPS bound. The no-force condition is satified
for negative charges because the timelike
black hole singularity
has negative energy \homa\ (making it difficult to model
with D-branes).}
If we further define
\eqn\fdf{Z_n \equiv -1+{r_n^2 \over r^2}}
in the interior region one finds that
the anti-one-brane moduli space metric is given by \onemod\
with the redefined $Z$s. Transforming to a new radial
coordinate $\rho={r_nr_5 \over r}$,
the interior moduli space metric is
\eqn\nmt{ds^2_{M1}=(1-{r_n^2 \over \rho^2})(1-{r_5^2 \over \rho^2})
(d\rho^2+\rho^2 d\Omega^2)+({ \rho^2\over r^2_5}-1)dx^mdx^m ,}
where now the horizon is at $\rho=\infty$.

\subsec{U-Dual Moduli Space Metric}

In this subsection we derive the manifestly U-dual expression which reproduces
\onemod\ and \wrfve\ for appropriate values of the charges.
The bosonic terms of the
low-energy effective action are manifestly U-dual in the
five-dimensional Einstein frame \sdd\
\eqn\uact{S={1 \over 4\pi^2 l_p^3}  \( \int d^5x\sqrt{-g}\[ R - {{\cal
G}_{\alpha\beta}\over 2}
\nabla_\mu \phi^\alpha \nabla^\mu \phi^\beta-
{{\cal M}_{IJ}(\phi )\over 4}
F^I_{\mu \nu}
F^{J\mu \nu} \]  +...
\) , }
where $\phi^\alpha, ~~\alpha=1,..42$ parametrizes an $E_{6,6}/USp(8)$ coset
with metric $\cal G$,
upper (lower) $I=1,...27$ is an index in the $27 ~~
(\bar{27})$ of $E_6$ and $F^I=dA^I$.
An extremal black hole
(probe) is
characterized by 27 electric charges which we shall denote $Q_I~~(q_I)$. The
black hole (probe) mass $M~~ (m)$ and scalar charges $\Sigma_\alpha
{}~~(\sigma_\alpha)$ are determined by supersymmetry in terms of these
27 electric charges. The long range $1/r^3$ force between the
black hole and probe is proportional to the products of the various charges.
This force vanishes when the probe is stationary and if, as in the cases we
consider, the black hole and probe charges preserve a common supersymmetry.
This implies that the static interaction energy vanishes
\eqn\srel
{{l_p^3 \over r^2}\( {mM \over 3}-q_I({\cal
M}^{-1})^{IJ}Q_J+\sigma_\alpha ({\cal G}^{-1})^{\alpha\beta}\Sigma_\beta
\) =0. }
The quadratic velocity-dependent interaction energy can be deduced from
the first quantized particle action
\eqn\partact
{ -\int d\tau \,\left\{ (m +\s_\alpha \phi^\alpha) \det\!^{1/2}(g_{\mu\nu} \dot
X^\mu
\dot X^\nu) + q_IA^I_\mu \dot X^\mu \right\}\ ,
}
with $\alpha$ an arbitrary constant.  Expanding to order $v^2$,
the first, gravitational+scalar, term in \srel\ is
multiplied by
$1- v^2$.\foot
{The naive $v^2/2$ from the kinetic energy is only half the
effect.  The other half is from the spatial metric, and depends on the
fact that $g_{ij} \sim (-g_{00})^{-1/2} \delta_{ij}$ for the
five-dimensional black hole Einstein metric at long distance.}
The electric term is uncorrected. Using the relation \srel\ between the
charges and the moduli, the $v^2/r^2$ interaction energy can be written
\eqn\eint{ {l_p^3 v^2\over 2r^2}\(mM-q_I({\cal M}^{-1})^{IJ}Q_J \).}

Next let us consider the $1/r^4$ terms in the moduli space metric.
These control the $r\rightarrow 0$ limit, which is the near-horizon geometry.
Thus they must be independent of the asymptotic moduli \fks.
There is only one moduli-independent invariant that
can be constructed from one probe and two black hole charges:
$d^{IJK}q_IQ_JQ_K$ where $d_{IJK}$ is proportional to 
the cubic $E_6$ invariant. 
The complete U-dual probe action is then
\eqn\prt{\int d\tau\,\left\{- m +{v^2 \over 2}\(m +{mM l_p^3-q_I({\cal
M}^{-1})^{IJ}Q_Jl_p^3\over r^2}+{d^{IJK}q_IQ_JQ_K l_p^3\over r^4}
\)\right\}.}

For example,
using the expression
\eqn\lpl{l_p=\({g^2 \apm^4 \over VR}\)^{1/3}}
for the five-dimensional Planck length in terms of the string conventions,
the one-brane probe action \onemod\ becomes
\eqn\efr{\int d\tau \,\left\{ -\biggl[{R^3 \over l_p^3
g^2V}\biggr]^{1/4} +{v^2 \over 2}
\( \biggl[{R^3 \over l_p^3 g^2V}\biggr]^{1/4} +\biggl({l_p^9 \over
g^2 RV}\biggr)^{1/4}{n
\over r^2} +{R Q_5 \over r^2}+{Q_5 n l_p^3 \over r^4}\) \right\}}
in the Einstein frame.
As expected, the $1/r^4$ term does not depend on the moduli.

\subsec{Geodesic motion}

Moduli space is a euclidean space and so a moduli space metric
cannot have an event horizon. Nevertheless we expect that the
probe can fall behind the black hole event horizon and this should
somehow show up in the moduli space metric.  The existence of an event
horizon shows up in the moduli space geodesics which describe the
motion of the probe. While \wrone\ and \wrfve\ are geodesically
complete, there are some geodesics which fall into the wormhole and
never return to the asymptotic region.\foot{The fact that these geodesics
take infinite time to reach the horizon at $r=0$ reflects our use of
Schwarzchild time in describing the motion.} These correspond to
probes which are captured by the black hole.

\def\P{r_n^2}
\def\Q{r_5^2}
To analyze the geodesic motion
consider an incoming one-brane with asymptotic velocity $v$ (and $w=0$)and
impact parameter $b$ (in units of $\apm$).
Motion in \wrone\ conserves the energy
\eqn\oneham{
E = {p^2\over 2Z_nZ_5} +
{L^2\over 2r^2 Z_nZ_5}
}
and angular momentum $L=r^2 Z_nZ_5\dot\theta$.
Here $p=Z_nZ_5\dot r$
is the canonical radial momentum.
The asymptotic values are $E=v^2/2$ and $L=bv$.
Solving for $p$ produces
\eqn\eqmotion{\eqalign{
{dr\over dt} &= {1 \over Z_nZ_5}\(2E Z_nZ_5-{L^2\over r^2}\)^{1/2} \cr
&= {v \over Z_nZ_5}\(Z_nZ_5-{b^2\over r^2}\)^{1/2}.
}}

The qualitative features of the motion can be understood
by finding the turning points $r_c$ at which $dr/ dt=0$.
These are at
\eqn\nbm{
r_c^2 = {b^2-\P-\Q\over 2} \pm
\half\sqrt{b^4-2b^2(\P+\Q)+(\P-\Q)^2}.}

In the simplified case $r_n=0$ (or $r_5=0$), there is a long tube but
no second asymptotic region.
For $b^2>\Q$, the turning point is real,
and the particle will re-emerge, after a time delay
$\pi\Q/r_c v$.
For $b\le r_5$, $\dot r$ never becomes zero, and
the incoming probe will approach $r=0$ monotonically.
At late times $r\ll r_5$ and the solution behaves as
$r \sim \exp -(2E\Q-L^2)^{1/2}t/\Q$.
Thus it is captured by the black hole.

More generally, the probe is captured if
$b \le b_c = r_n+r_5$.
The wormhole geometry \wrone\ has a minimal closed geodesic at
$r_c = \sqrt{r_n r_5}$ which influences the motion.
This can be seen by considering $b = b_c$.
Now
\eqn\rft{{dr\over dt} = \pm v
{1 - {r_n r_5\over r^2}
\over Z_nZ_5}.}
At this critical impact parameter,
the two real solutions of $\dot r = 0$ coalesce at $r_c$.
Near $r_c$ the motion is $\dot r \sim -v c r$ and the trajectory asymptotes
to $r_c$.
For $b<b_c$, the turning points move off the real axis.
The probe is slowed down as it passes
$r_c$, but will eventually asymptote to $r=0$.

\newsec{Static Probes of a Near-Extremal Black Hole}

Supersymmetry implies that all configurations consisitng of a static
one-brane probe plus an extremal black hole have degenerate energies.
This degeneracy is lifted when supersymmetry is broken by excitng the probe.
In the previous section the corresponding velocity and position dependent
action was computed. In this section we shall consider the closely related
problem in which the black hole rather than the probe is excited.  The black
hole then becomes near-extremal,
the long range forces on a static probe no longer exactly cancel and the probe
action acquires a potential term. We shall see that this potential has
a structure similar to the metric of the preceding section.

\subsec{Relevant Formulae for Near-Extremal Black Holes}
The relevant formulae were collected in
\refs{\hms,\msg} from which most of this section was taken.  In this section 
we set $\apm = 1$. We will work with the following 
near-extremal solution
labeled by three charges \hms ,
given in terms of the ten-dimensional
variables by
\eqn\ndil{ e^{-2\phi } = {Z_5 \over Z_1},}
\eqn\nhtn{H= 2 r_5^2 \e_3 + 2 r_1^2  e^{-2\phi}  *_6 \e_3,}
\eqn\nmetric{\eqalign{
ds^2 = &Z_1^{-1/2}
Z_5^{-1/2}
  \[ - dt^2 +dx_5^2+{r_0^2  \over r^2} (\cosh\sigma dt + \sinh \sigma
dx_5)^2\]
\cr
+&
 Z_1^{1/2}
Z_5^{-1/2}dx^m dx^m  +Z_1^{1/2} Z_5^{1/2}\left[
Z_0^{-1} dr^2 + r^2 d \Omega_3^2 \right]
,}}
where in this section we define
\eqn\ndfr{\eqalign{Z_1&=1+{r_1^2 \over r^2},\cr
Z_5&=1+{r_5^2 \over r^2},\cr
Z_0&=1-{r_0^2 \over r^2},\cr}}
with
\eqn\paramsol{\eqalign{
r_1^2 &=\sqrt{\({g Q_1  \over V }\)^2+{r_0^4 \over 4}}~-{r_0^2 \over 2} ,\cr
{}~~~~~~
r_5^2 &=\sqrt{{(g Q_5 ) }^2+{r_0^4 \over 4}}~-{r_0^2 \over 2},\cr
{}~~~~~~r_0^2 {\sinh 2  \sigma \over 2}
&= { g^2 n \over R^2 V },\cr~~~~~~
r_n^2 & \equiv r_0^2 \sinh^2  \sigma, }}
The extremal limit is $r_0 \to 0, ~\sigma \to \infty$ with
$n$ held fixed.

In the dilute gas region\foot{
This corresponds to the limit $\alpha, \gamma \gg \sigma $
of the solution in \hms, which is the dilute gas region discussed
in \ghas \hms \msg.} defined by
\eqn\smallp{
r_0 , r_n \ll r_1 , r_5 ,
}
the energy is approximately
\eqn\extl{\eqalign{E &
= {1 \over g^2}\left[  {R gQ_1}  +
{  R V gQ_5}+{g^2n \over R} +{VRr_0^2e^{-2\s}\over 2}\right]   ~.\cr}}

The momentum $n$ is carried by
a gas of left and right movers on the effective string. Equating the energy of
this
gas to $ {n \over R }+{RVr_0^2e^{-2\s } \over 2g^2}$ and its
momentum to ${n\over R}$ we can determine $n_L$ and $n_R$ :
\eqn\nlr{\eqalign{n_L&= {n  }+{R^2Vr_0^2e^{-2\s } \over 4g^2},\cr
                  n_R&= {R^2Vr_0^2e^{-2\s } \over 4g^2}.\cr} }
The left and
right
moving oscillations are governed by  effective left and right moving
temperatures
\eqn\tleft{\eqalign{
T_L &= {1\over \pi } {r_0 e^\sigma  \over 2 r_1 r_5 }
={g\over \pi R r_1r_5}\sqrt{n_L \over V}, \cr~~~~~~~~~~~
T_R &= {1\over \pi } {r_0 e^{-\sigma}  \over 2 r_1 r_5 }
={g\over \pi R r_1r_5}\sqrt{n_R \over V}. ~~~~~~
}}
Other useful relations are
\eqn\rz{r_0^2={4g^2\sqrt{n_Ln_R}\over R^2V}=4\pi^2T_LT_Rr_1^2r_5^2,}
\eqn\sma{e^{2\s}=\sqrt{n_L\over n_R}.}

\subsec{The Probe Action}

The action for a one-brane probe is
\eqn\news{
-{1\over 2\pi g}\int d\tau
 d\sigma\ e^{-\phi}\sqrt{-\det g_{\mu\nu}\p X^\mu \p X^\nu}
+ {1 \over 2\pi g} \int C^{(2)},
}
where
\eqn\ctew{C^{(2)}= {gQ_1dt\wedge dx^5 \over r^2 VZ_1}}
is the potential obeying $H=dC^{(2)}$.
For a static probe we take
\eqn\onns{\eqalign{X^0&=\tau , \cr X^5&=\sigma, \cr
                  X^i&=constant, \cr X^m &=constant~.}}
For a static probe in an extremal black hole geometry there is an
exact cancellation between the two contributions
to the action. Hence there is no potential and no force on the
probe. This cancellation no longer occurs when the black hole is
excited.
The first term is \eqn\rrf{-\int d\tau{R\sqrt{Z_0}\over g Z_1}}
while the second is
\eqn\ssd{-\int d\tau{RQ_1 \over r^2 V Z_1},}
so the total action is
\eqn\ts{S=-\int d\tau{R \over g}{\sqrt{Z_0}+{gQ_1 \over r^2V} \over
Z_1}
\equiv -\int d\tau({R \over g}+U).}
We wish to expand this in $r_0$. Note that
\eqn\zex{\sqrt{Z_0}=1-{r_0^2 \over 2 r^2}-{r_0^4\over 8 r^4}+...}
\eqn\zoex{Z_1=1+{gQ_1 \over r^2 V}-{r_0^2 \over 2 r^2}
+{r_0^4\over 8 r^2 r_1^2}+...}
To leading order in $r_0$ the potential energy is then
\eqn\cfg{U=- {Rr_0^4 \over 8gr^2r_1^2}
=-{2g^2{n_Ln_R}\over R^3Vr^2Q_1}=-{2\pi^4Rg^2Q_1Q_5^2 T_L^2T_R^2
\over r^2 V}.}
Let us now consider the case $T_L\gg T_R$ so that $n_L \sim n$ and
$T_R \sim T_H/2$, where $T_H$ is the Hawking temperature. Then
\eqn\usin{U \sim -{\pi^2 g^2 nQ_5 T_H^2 \over 2 r^2RV}.}
We have been working in string units. Transforming to
five-dimensional Planck units using
\eqn\spl{\alpha^\prime =({l_p^3 VR\over g^2})^{1/4},}
the extra term in the action becomes
\eqn\sbc{\Delta S =\int d \tau {\pi^2 n Q_5 T_H^2 l_p^3\over 2 r^2
}.} Note that all moduli dependence has disappeared and the structure is
similar to the $1/r^4$ term in the moduli space metric.  It would be
interesting to
reproduce this term from D-brane perturbation theory.

%

%
\def\IZ{\relax\ifmmode\mathchoice
{\hbox{\cmss Z\kern-.4em Z}}{\hbox{\cmss Z\kern-.4em Z}}
{\lower.9pt\hbox{\cmsss Z\kern-.4em Z}}
{\lower1.2pt\hbox{\cmsss Z\kern-.4em Z}}\else{\cmss Z\kern-.4em
Z}\fi}
\def\IB{\relax{\rm I\kern-.18em B}}
\def\IC{{\relax\hbox{$\inbar\kern-.3em{\rm C}$}}}
\def\ID{\relax{\rm I\kern-.18em D}}
\def\IE{\relax{\rm I\kern-.18em E}}
\def\IF{\relax{\rm I\kern-.18em F}}
\def\IG{\relax\hbox{$\inbar\kern-.3em{\rm G}$}}
\def\IGa{\relax\hbox{${\rm I}\kern-.18em\Gamma$}}
\def\IH{\relax{\rm I\kern-.18em H}}
\def\II{\relax{\rm I\kern-.18em I}}
\def\IK{\relax{\rm I\kern-.18em K}}
\def\IP{\relax{\rm I\kern-.18em P}}
\def\IR{\relax{\rm I\kern-.18em R}}

%

%
\def\II{\relax{I\kern-.10em I}}

\def\IIb{{\II}b}
\def\e{\epsilon}
\def\apm{{\alpha^\prime}}

\def\[{\left [}
\def\]{\right ]}
\def\({\left (}
\def\){\right )}
\def\p{\partial}
\def\vol{{\rm vol~}}

\lref\juanthesis{
J. M. Maldacena, ``Black Holes in String Theory,'', Ph.D. Thesis.,
hep-th/9607235.}
\lref\insta{
E. Witten, Nucl. Phys. {\bf B460} (1996) 541, hep-th/9511030.}
\lref\instb{
M. R. Douglas, ``Branes within Branes,'' hep-th/9512077.}
\lref\instc{
C. Vafa, Nucl. Phys. {\bf B463} (1996) 435, hep-th/9512078.}
\lref\ghr{S. J. Gates, C. M. Hull, and M. Ro\v cek, Nucl. Phys. {\bf
B248} (1984) 157.}

\newsec{Low Energy D-Brane Field Theory}

We now consider the same system in the regime of D-brane perturbation
theory.  The black hole consists of $Q_1$ D one-branes parallel to the
5-direction and $Q_5$ D five-branes parallel to the 56789-directions,
carrying 5-momentum $P$.
The D one-brane probe is at a distance $r$
small compared to the string scale.  In this regime the relevant states
are open strings that are massless or become massless as $r \to 0$.
We will denote the string endpoints by 1, 5, and $1^*$, the latter
referring to endpoints on the probe.

The massless $1^* 1^*$ states are a gauge field $B^\mu$, a collective
coordinate $X^i$ (recall $i = 1,2,3,4$), and their fermionic partners,
together forming a vector multiplet, and a collective coordinate $X^m$
($m = 6,7,8,9$) and its fermionic partners in a hypermultiplet.
We are primarily concerned here with the effective action for $X^i$.

On the black hole are $Q_1^2 + Q_5^2 + Q_1 Q_5$ hypermultiplets,
of which $Q_1^2 + Q_5^2$ receive mass from the $U(Q_1)$ and $U(Q_5)$
D-terms \juanthesis.  It would thus appear to be natural to
represent the moduli as 15 open string fields.  However, this description
is simple only near the point in moduli space where these fields vanish.
When these fields are large, the one-branes dissolve into the
five-branes, becoming $Q_1$ $U(Q_5)$ instantons
\refs{\insta,\instb,\instc}.  This instanton moduli space gives a global
description of the black hole hypermultiplet moduli space.
That is, on the five-branes is a self-dual gauge field
$A_m(x^n,\zeta)$, where $\zeta$ are $4Q_1 Q_5$ parameters.  The low
energy fields are given by letting $\zeta$ depend on $x^\mu$,
$\mu = 0,5$.
There are in addition $4Q_5^2$ scalars $Y^i$ in the 55 vector multiplet.
However, generically the instanton gas breaks the $U(Q_5)$ down to
$U(1)$,\foot
{For a static configuration ($n=0$), this is true only when
$Q_1 \geq Q_5$.  But when $n$ is macroscopic it seems likely that
the fluctuations fill out and break the full group.} so only the center
of mass part of $Y^i$ is massless.

The interaction between the probe and the black hole comes from $1^* 5$ strings
with one end on each.  Note that there are no
$1^* 1$ strings: we are in the regime where the one-branes are
described by five-brane gauge fields, not by D one-branes.  The $1^* 5$
strings have mass proportional to $r$.  Terms in the effective
action which are singular in $r$ are obtained by integrating out the
virtual $1^* 5$ strings.  These fields comprise $Q_5$ hypermultiplets,
with bosonic components
$\phi^{A'a}$ and fermionic components $\chi^{Aa}$.  These are complex
fields, with $a$ a $U(Q_5)$ index.  The indices $A$ and $A'$ are defined
as follows.  Call
$SO(4)_E$ the rotational symmetry group of transverse space, and
$SO(4)_I$ the local Lorentz symmetry in
$T^4$.  We index doublets of the two $SU(2)$'s in $SO(4)_E$ as $A$ and
$Y$, and doublets of the two $SU(2)$'s in $SO(4)_I$ as $A'$ and $\tilde
A'$.

To simplify the discussion we will take as
external fields only the massless bosonic fields, namely $B_\mu$, $X^i$,
and $X^n$, the self-dual part of $A_m$, and the $U(1)$ parts of
$A_\mu$ and $Y^i$.  The full
$1^* 5$ hypermultiplets run in the loop.
The minimally coupled action is
\eqn\lowact{\eqalign{
S_0 = & -{\mu\over g}\int d^2x \( 1 + {1\over 2}\p_\mu X^i \p^\mu
X^i + {1 \over 4\mu^2} G_{\mu\nu} G^{\mu\nu} + {1\over 2} \p_\mu X^m
\p^\mu X^m \right.\cr
& + D_\mu \phi^\dagger D^\mu \phi  + \mu^2 \(X^{i2}
\phi^\dagger\phi - 2 X^{i}
\phi^\dagger Y^i \phi +
\phi^\dagger Y^{i2}\phi \) \cr
&\left. + {\mu^2 \over 8} 
 \bigl| \phi^{\dagger A'a}\phi^{B'}_a +
\phi^{\dagger A'a}\phi^{B'}_a \bigr|^2
+ \bar\chi (\Gamma^\mu D_\mu + i\mu \Gamma^i X^i) \chi \)\cr
& -{\mu^3 \over 4 \pi^2 g}\int d^6x\, {\rm Tr}\(
1 + {1\over 2}\p_M Y^i \p^M Y^i +
{1\over 4\mu^2 } F_{MN} F^{MN}\)\ .}}
Here $\mu$ is the fundamental string tension
$(2\pi\apm)^{-1}$; $G_{\mu\nu}$ is the
$1^*1^*$ field strength and
$F_{MN}$ the $55$ field strength, with $M$ running over $\mu$
and $m$.
The quartic potential terms can be understood repectively from the
dimensional reduction of the $d=6$ covariant derivative and from the
$1^*1^*$ and $55$ $U(1)$ $D$-terms.  The $SU(Q_5)$ $D$-terms must be
absent (cancelled by exchange of massive fields) because this group is
broken. There is also a  five-brane $U(1)$ $D$-term which is omitted
from \lowact\ because it is proportional to $1/V$ and can be
suppressed by large $V$. 

The action for the black hole moduli $\zeta$ comes from the gauge
kinetic term $F_{MN} F^{MN}$, specifically the components $F_{m\mu}
F^{m\mu}$.  In using the low energy effective action for the these
gauge fields we are treating the instantons as large compared to the
string scale.  Effectively this is the first term in an expansion in
$1/V$, since as the volume is scaled up the gauge fields scale
uniformly.  However, we will see that only one coupling matters,
involving a conserved charge on the black hole side, and it is likely
that its coefficient is universal.

In the minimal action~\lowact\ there is no coupling between the black
hole moduli and the other fields.  However, there are non-minimal
(Born-Infeld-like) couplings that are important.  Although nominally
higher derivative, they contribute at leading order because of the
nonzero 5-momentum of the black hole.  To be somewhat systematic, note
first that the fields and couplings appear in the D-brane
action~\lowact\ and the expected moduli space metric~\wrone\
only in the combinations
$$\matrix{\apm^{-1} g :\ m^2& \apm^{-1} X^i ,\ \apm^{-1} X^m,\ B^{\mu},\ A^\mu,\
\apm^{-1} \phi:\ m \cr
\apm^{-1} \chi :\ m^{3/2} & \apm A_m:\ m^{-1}\qquad \apm^{-4} V:\
m^{4} }$$
with units as shown.  Any terms accompanied by explicit powers of
$\apm$ in addition to these will be suppressed at low energy.
This
unusual dimensional analysis can be deduced by starting with the
observation that
the probe-black hole separation~$X^i$ enters only through the masses of the
$1^*5$
strings.
It allows at the same order as the action~\lowact\ a number of new
terms; the one relevant for the present purposes is
$F_{-m} F_{-m} D_+\phi^\dagger D_+ \phi$, through which the $1^* 5$ strings
couple
to the momentum carried by the black hole.

We can find this term and determine its coefficient by a $T$-duality
argument, just
as for the Born-Infeld action (see~\dbr\ for a review with references).
Let us first find
terms with $\p_- X^m$.  To
do this consider a one-brane probe boosted so that $X^m$ is linear in
$x^{-}$,
\eqn\tilt{X^m = u^m x^- .}
In the rest frame of the probe the action is~\lowact.
The rest frame (primed) coordinates
are related to the lab (unprimed) by
\eqn\boost{x'^- = x^-,\qquad x'^+ = x^+ - 2 u_m x^m + u^2 x^-, \qquad
x'^m = x^m - u^m x^- . }
The spinor indices should be thought of as tangent space indices,
so they do not transform.
For $1^* 5$ fields, which live on $x'^m = 0$, the derivatives are
related
\eqn\boostder{
\p'_+ = \p_+, \qquad \p'_- = \p_- + u^2 \p_+.
}
In the lab frame the action then has the additional terms
\eqn\moreact{
\delta_1 S =  -{\mu \over g}\int d^2x\,  2 \p_- X^m \p_- X^m \( \p_+
X^i
\p_+ X^i + 2 D_+ \phi^\dagger D_+ \phi
+ \bar\chi \Gamma_+ D_+ \chi \).}
In addition, $A_-$ picks up a non-gauge piece $A_m \p_- X^m$.
Terms involving $F_{-m}$ are now obtained simply by $T$-duality.  This
takes $X_m
\leftrightarrow \mu^{-1} A_m$ and gives
\eqn\evenmore{\eqalign{
\delta_2 S = & -{1\over g\mu}\int d^2x\,  2 {\rm Tr}(F_{-m} F_{-m}) \( 2 D_+
\phi^\dagger D_+ \phi  + \bar\chi \Gamma_+ D_+ \chi \)\cr&
-{\mu^3 \over 4 \pi^2 g}\int d^6x\, {\rm Tr}\( 2 F_{-m} F_{-m} \p_+ Y^i \p_+
Y^i\) .}}
The modification of $A_-$ is self-dual, up to charge conjugation
and a gauge transformation.

The momentum carried by the black hole is
\eqn\noether{
{n\over R} = (2\pi)^5 R V T^{05} = {(2\pi)^3 R V \mu \over  g}{\rm
Tr}(F_{-m}^2).
}
Hence $u^2$ maps to
\eqn\utwo{
\tilde u^2 = { g n\over (2\pi)^3 R^2 V \mu^3 Q_5},
}
with the assumption that $F_{-m}^2$ is on the average
proportional to the identity.
Collecting only the terms that survive in the low energy limit of
interest gives the action
\eqn\fullact{\eqalign{
S = & -{\mu\over g}\int d^2x \( {1\over 2}\p_\mu X^i \p^\mu
X^i + {1 \over 4\mu^2} G_{\mu\nu} G^{\mu\nu} + {1\over 2} \p_\mu X^m
\p^\mu X^m \right.\cr
& + \tilde D_\mu \phi^\dagger \tilde D^\mu \phi  + \mu^2 X^{i2}
\phi^\dagger\phi
+ {\mu^2 \over 8} \bigl| \phi^{\dagger A'a}\phi^{B'}_a +
\phi^{\dagger A'a}\phi^{B'}_a \bigr|^2 \cr
&\left. + \bar\chi (\Gamma^\mu D_\mu + i\mu \Gamma^i X^i) \chi
+ {2 \over \mu^2} {\rm Tr}(F_{-m} F_{-m}) \( 2 D_+
\phi^\dagger D_+ \phi  + \bar\chi \Gamma_+ D_+ \chi \) \)
\cr
& +{\mu^3 \over 4 \pi^2 g}\int d^6x\, {1\over 2\mu^2 }{\rm Tr}\(
F_{-m} F_{+m}\).}}

Finally we can do the one loop graph $1^*5$ graph.  First go to the
one-plus-one dimensional limit, $R \to \infty$, giving
\eqn\dets{\eqalign{&
Q_5 \log\det\(\Gamma^\mu \p'_\mu + i \mu \Gamma^i X^i\) - 2Q_5\log\det
\(\p'^\mu \p'_\mu - \mu^2 X^{i2} \) \cr
=\ & {Q_5 \over 2}\log\det\(\p'^\mu \p'_\mu - \mu^2 X^{i2} + i \mu
\Gamma^\mu
\Gamma^i
\p'_\mu X^i\) - 2 Q_5 \log\det \(\p'^\mu \p'_\mu - \mu^2 X^{2} \).
}}
Here we have defined
\eqn\ders{
\p'_+ = \p_+, \qquad \p'_- = \p_- + \tilde u^2 \p_+.
}
The only external $1^*1^*$ fields that have been kept are the $X^i$.  The
Dirac matrices are four-dimensional.  The first nontrivial term is quadratic
in the velocity, giving
\eqn\oneloop{
-\mu^2 Q_5 \p'^\mu X^i \p'_\mu X^i \int {d^2 p\over (2\pi)^2}
(p'^\mu p'_\mu + \mu^2 X^{2} )^{-2}
= -{i Q_5\over 4\pi}(1 + \tilde u^2) {\dot X^2 \over X^2 } .
}
The zero plus one loop effective action for the probe is then
\eqn\metric{
-{\mu\over2g} \dot X^2 \(1 + {Q_5 g\over \mu X^2} +
{g^2 n\over R^2 V \mu^4 X^2} \)\ ,
}
the same as the metric~\wrone\ from the supergravity regime, to this
order.

For $R$ finite one can make the usual long-string argument \mss.
The effective length of the string is $2\pi R Q_1 Q_5$.  In the
classical limit this is long compared to Compton wavelength of the $1^*
5$ strings and so the above calculation still holds.

At two loops, where we might expect to see the full metric~\wrone,
there is a puzzle.  There appear to be no two-loop graphs having
$r^{-4}$ singularities.  All 55 fields are massive, as we have
discussed, and give rise to low energy interactions that are higher
order in the above dimensional analysis.  Interactions involving $1^*
1^*$ fields cannot contribute.  If the $1^* 1^*$ field is inside a loop
the graph is suppressed by large-$N$ counting in the classical limit.
Otherwise, the graph must be one-particle-reducible with respect to the
$1^* 1^*$ field and so represents mixing (e.g. $X^i X^\mu$), but this
is forbidden by symmetries.  There remain only the quartic
interactions among the $1^* 5$ fields (which, in superfield language,
are actually 1PR on the $1^* 1^*$ superfield).  These give a
figure-eight graph, with a 5-index running inside each loop and the
$1^*$ around the outside, and with the external fields attached at
various points. Now, both
$F_{-m}$ must attach to the same loop in order to give Tr$(F_{-m})^2$.
Lorentz invariance then requires both $\p_+ X^i$ on that same loop, but
then the other loop has no external attachments and vanishes by
supersymmetry.

It is possible that the two loop term cannot be seen in the
framework~\lowact.  We justified this expansion by considering a
large-$V$ limit where the generic gauge field becomes smooth.
It may be that special points in the instanton moduli space,
such as zero-size instantons, are the source of the missing term.  This
is under investigation.

\newsec{Constraints from Supersymmetry}

In this section we consider the constraints of supersymmetry on the
moduli space metric.  They are consistent with the results of the
previous section.

First consider the limit $n=0$, where the low energy theory has $(4,4)$
supersymmetry.  The coordinates $X^i$ are in an $(4,4)$ twisted
chiral (=vector) multiplet~\ghr.  This can be written in terms of two
$(2,2)$ superfields, $\phi$ being chiral and $\chi$ twisted chiral.  These
satisfy
\eqn\chiralcon{\eqalign{
&\bar D_+ \phi = \bar D_- \phi = 0 \cr
&\bar D_+ \chi = D_- \chi = 0.
}}
Their lowest components are
\eqn\comps{
\phi|_{\theta = 0} = X^3 + i X^4, \qquad \chi|_{\theta=0} = X^1 + iX^2.
}
The $N=1$ action
\eqn\susact{
S_1 = \int
d^2x\,d\theta_+\,d\theta_-\,d\bar\theta_+\,d\bar\theta_-\,
K(\phi,\bar\phi,\chi,\bar\chi)
}
gives the metric
\eqn\metmod{
ds^2 = K,_{\phi\bar\phi} d\phi d\bar\phi -
K,_{\chi\bar\chi} d\chi d\bar\chi.
}
The condition for $(4,4)$ supersymmetry is~\ghr
\eqn\fourfour{
K,_{\phi\bar\phi} + K,_{\chi\bar\chi} = 0,
}
which is simply Laplace's equation.  The function $K$ is not observable
and need not be spherically symmetric, but Laplace's equation also follows
for the metric compenents themselves and these must be spherically
symmetric.  Thus
\eqn\twomet{
g_{\phi\bar\phi} = g_{\chi\bar\chi} = a + {b\over X^2}.
}
It follows that the dependence at $r \to \infty$ determines that at $r \to
0$.  Similarly the $Q_5 = 0$ metric must be of the same form.

Now look at $Q_5$ and $n$ both nonzero.  The unbroken supersymmetry is
$(4,0)$.  A $(4,0)$ invariant operator with two lower $-$ indices
must multiply $n$ in the low energy theory.\foot
{It is conceivable that the full $(4,4)$ invariance will
give a stronger constraint.}  This is easily written down with a $(4,0)$
superfield.  Let us review the $(4,4)$ superfield for the twisted
chiral multiplet~\ghr.  It consists of two functions $\hat\phi,
\hat\chi$, of $\theta_{a\pm}$ and their conjugates, $a = 1,2$.  These
satisfy differential constraints that completely determine their
dependence on $\theta_{2\pm}, \bar\theta_{2\pm}$, so one can express
everything in terms of the values at $\theta_{2\pm} = \bar\theta_{2\pm} =
0$; these are just the unhatted $(2,2)$ superfields above, with $\theta_{1}
\to\theta$.

The same works for $(4,0)$ with just the $\theta_{a+}$.
The action is
\eqn\sustwo{
S_2 = n\int
d^2x\,d\theta_{1+}\,d\bar{\theta}_{1+}\,d\theta_{2+}\,d\bar{\theta}_{2+}\,
F(\hat\phi,\bar{\hat\phi},\hat\chi,\bar{\hat\chi})
}
and the constraints are~\ghr
\eqn\chiralcontwo{\eqalign{
&\bar D_{a+} \hat\phi = \bar D_{a+} \hat\phi = 0 \cr
& (D_{2+} + i\bar D_{1+}) (\bar\phi + \chi) = 0 \cr
& (D_{2+} - i\bar D_{1+}) (\phi + \bar\chi) = 0 .
}}
Using these constraint one can reduce to $(2,2)$ superfields,
\eqn\twotoone{\eqalign{
\int
d\theta_{2+}\,d\bar{\theta}_{2+}\,
F
\sim\ & D_{2+}\,\bar D_{2+}\,
F\bigr|_{\theta_{2+}
=\bar\theta_{2+} = 0}\cr
=& \(F,_{\phi\bar\phi} + F,_{\chi\bar\chi}\)\(D_{1+}\phi \bar
D_{1+}\bar\phi + D_{1+}\chi \bar D_{1+}\bar\chi\)|_{\theta_{2+}
=\bar\theta_{2+} = 0}\ .
}}
The bosonic part of the action is then
\eqn\bos{
\nabla^2 F n \p_+ X^i \p_+ X^i .
}
A general spherically symmetric $F$ allows a general $r$-dependence.

\centerline{Acknowledgments}
We are grateful to T. Banks, S. Shenker for collaboration
at earlier stages of this work, and discussions at all stages. 
We would also like to thank G. Horowitz, J. Maldacena 
and L. Susskind for discussions. Results related 
to those of Section 2 appeared recently in 
\nref\tset{A. Tseytlin, hep-th/9702163.} \tset. This work was supported 
in part by DOE grants 
DOE91ER-40618 and DE-FG02-96ER40559 and NSF grant PHY94-07194. 
\listrefs
\end